\documentclass[a4paper]{llncs}

\usepackage{local}

\title{Metrics and Ambits and Sprawls, Oh My}
\subtitle{Another Tutorial on Metric Indexing}
\author{Magnus Lie Hetland}
\institute{Norwegian University of Science and Technology,
\email{mlh@ntnu.no}}

\begin{document}
\maketitle

\begin{abstract}
    A follow-up to my previous tutorial on metric indexing, this paper walks
    through the classic structures, placing them all in the context of the
    recently proposed \emph{sprawl of ambits} framework.
    The indexes are presented as configurations of a single, more general
    structure, all queried using the same search procedure.
\end{abstract}

\section{Introduction}

About ten years ago, I wrote a tutorial on metric
indexing~\cite{Hetland:2009}, and last year, I finally finished a unifying
framework for metric indexing and other comparison-based
indexing~\cite{Hetland:2019}. That paper, however, is perhaps not the most
inviting, containing quite a bit of detail and formalism, so in this paper,
I'll revisit my earlier tutorial, in light of this new framework. This
approach has two main benefits. First, the result should ideally be a
streamlined, unified tutorial, rather than a smorgasbord of disjoint
techniques; and, second, it provides an example-based introduction to
the framework of sprawls and ambits, which might be useful to researchers who
are already familiar with metric indexing. I focus primarily on ``the
classics''; for an overview of many variants, see, e.g., the recent paper by
Chen et al.~\cite{Chen:2020}.

In contrast to the full paper introducing sprawls and
ambits~\cite{Hetland:2019}, I will try to keep this tutorial brief and to the
point---more so, even, than my previous tutorial. In keeping with that, let's
get going!

\section{Framework}

This section presents a thumbnail sketch of the framework used throughout. It
may be easier to understand \emph{after} you've read some of the example
applications, so feel free to skim it, then skip ahead to
\cref{sec:balltrees}, returning here later.

We have a data set $V$ drawn from some universe $U$ with an associated
\emph{metric}, i.e., a symmetric function $\delta:U\times
U\to\mathds{R}_{\geq 0}$, where $\delta(u,v)=0$ iff $u=v$, and
\begin{equation}
    \delta(u,v) \leq \delta(u,w) + \delta(w,v)\eqcomma
\end{equation}
for all $u,v,w\in U$. The problem we are trying to solve is storing $V$ in
some data structure, which we may later traverse to efficiently extract those
points relevant to some query. Intuitively, we view this data structure as a
bipartite digraph of points and \emph{regions}, i.e., \emph{sets} of
points. This is referred to as a \emph{sprawl} of
regions.\footnote{Equivalently, a hyperdigraph on $V$, with one region per
hyperedge~%
\cite[Rem.\,2.3.12]{Hetland:2019}.}

A region $R$ with parents $p_1,\dots,p_m$ is then \emph{defined} in terms of
these. That is, whether $u\in R$ depends on the distances
$x=[\delta(u,p_1),\dots,\delta(u,p_m)]$, a vector in the so-called \emph{pivot
space} of $p_1,\dots,p_m$.
Specifically, we use a linear function $f(x)$ and a threshold, or
\emph{radius} $r$, so $u\in R$ iff $f(x)\leq r$. Such a region is called a
linear \emph{ambit}.

The regions partition the space, representing a coarsening of the data. For a
query in the form of a ball $Q=\set{u:\delta(q,u)\leq s}$ of relevant points,
we are only interested in the contents of a region $R$ \emph{if it intersects}
$Q$. The idea, then, is to have the children of $R$ point the way to smaller
subsets of the data set. Search becomes a traversal of our graph, where each
region is checked for overlap with the query before possibly traversing its
children.

What is more, because a region is defined by its parents, we require
\emph{all} the parents to be traversed before traversing the region, and
possibly its children. When we traverse a point $u$, we compute $\delta(q,u)$,
so that when we traverse a region, we have all of
$\delta(q,p_1),\dots,\delta(q,p_m)$ available, giving us a distance vector $z$
representing the query. If we assume, for now, that $f$ is nondecreasing, $Q$
and $R$ intersect only if:\footnote{Here $f(s)$ is shorthand for
$f(s,\dots,s)$.}
\begin{equation}
    \label{eq:overlap}
    f(z) \leq r + f(s)
\end{equation}
\def\Queue{\gamma}%
For more advanced queries ($k$NN), and when we permit elimination, the order
of traversal is significant. In these cases, we'd use a priority queue of
nodes to traverse, updating their priorities each time we encounter them. In
the basic scenario sketched out here, though, we might as well use a
depth-first approach, as in the following mutually recursive procedures:

\bigskip
\noindent
{\tabcolsep=0pt
\begin{tabularx}{\textwidth}{p{.5\textwidth}p{.5\textwidth}}%
\toprule
Simplified sprawl search algorithm \\
\midrule
\begin{pseudo}[compact]*
\hd{Visit-Point}(u, q, s) \\
if $\delta(q,u)\leq s$ \label{l:memo} \\+
    print $u$ \\-
for $R\in u.\id{children}$ \\+
    $R.\id{count} = R.\id{count} + 1$ \\
    if $R.\id{count} \== |\mkern1mu R.\id{parents}\mkern1mu|$ \\+
        \pr{Visit-Region}(R, s) \\--
$u.\id{color} = \cn{black}$
\end{pseudo}%
&%
\begin{pseudo}[compact]*
\hd{Visit-Region}(R, q, s) \\
\nf get $z$ from $R.\id{parents}$ \\
\nf get $f$ and $r$ from $R$ \\
if $f(z) > r + f(s)$ \\+
    return \\-
for $v\in R.\id{children}$ \\+
    if $v.\id{color} \== \cn{white}$ \\+
        \pr{Visit-Point}(v, q, s)
\end{pseudo} \\
\bottomrule
\end{tabularx}

\bigskip
\noindent
In general, the idea is that $\delta$ is memoized in some way, so once
$\delta(q,u)$ is computed on line~\ref{l:memo} of \pr{Visit-Point}, it is
subsequently available when we gather up $z$ in \pr{Visit-Region}. Normally,
one would have one or more designated root nodes, and call \pr{Visit-Point} on
them in turn to initiate  the search.

The way this is set up, one would need to run a reinitialization in-between
queries, resetting the memo, coloring nodes white and setting counts to zero.
There are many ways of handling this, of course. One could have actual
attributes in the nodes, and maint a list of those that need resetting,
requiring constant amortized time. An even simpler approach might be to simply
use hash tables that are reset between searches. With some additional memory,
one could even do the reset in actual constant time, using the standard trick
for constant-time array initialization. In this case, one could keep a stack
of nodes whose attributes are valid, and let each node keep its index in the
stack. Then the reset would simply require setting the stack length to zero.

\section{Ball Trees}
\label{sec:balltrees}

A metric ball tree is a form of search tree where subtrees and their points
are enclosed in balls. A subtree is then only explored if its ball intersects
the query. For example, the simple BS-tree is a tree where each node is
associated with a single point and a radius that covers the points below it in
the tree~\cite{Kalantari:1983}. The idea of a sprawl is for the graph (in
this case, a tree) to express dependencies, where we have edges from points to
the regions they tell us about, and from regions to the points they tell us to
explore (\emph{if} we intersect them). In the case of the BS-tree, then, each
BS-tree node would be split into two sprawl nodes: one for the point, and one
for the radius (i.e., region). For example:

\bigskip
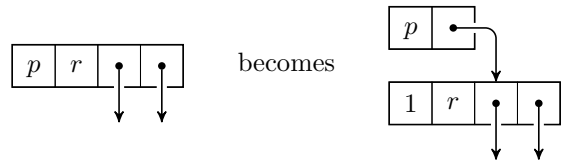
\begin{center}
\begin{tikzpicture}[baseline=-.5ex]

\matrix[memblock] (BS) { p \& r \& \& \\};

\draw
    (BS-1-3) node[tail] (BS-left) {}
    (BS-1-4) node[tail] (BS-right) {}
;

\draw[pointer] (BS-left) -- +(0,-.75);
\draw[pointer] (BS-right) -- +(0,-.75);

\end{tikzpicture}%
\qquad becomes\qquad%
\begin{tikzpicture}[baseline=-.5cm-.5ex]

\matrix[memblock, anchor=west] (P) { p \& \\};

\matrix[memblock, anchor=west] (R) at (0,-1) {
|[text depth=0]| 1
\& r \& \& \\};

\draw
    (P-1-2) node[tail] (P-ptr) {}
    (R-1-3) node[tail] (R-left) {}
    (R-1-4) node[tail] (R-right) {}
;

\draw[pointer] (P-ptr) -| (R-1-3);
\draw[pointer] (R-left) -- +(0,-.75);
\draw[pointer] (R-right) -- +(0,-.75);

\end{tikzpicture}
\end{center}
\bigskip

\noindent
Handling a BS-node then means first computing $\delta(q,p)$ and considering $p$ for
inclusion in the result, and then determining whether $\delta(q,p)$ is greater
than $r + s$. If so, no further action is taken, as the query ball $Q$ does
not overlap the region (i.e., ball) $R$. Otherwise, the two child pointers are
followed recursively.

In the sprawl version, we've split out the point $p$ as a parent node of the
region. Initially, we visit this node, compute $z=\delta(q,p)$, and increment
a counter associated with the child node. In general, we'll need to hang on to
the $z$ value as well as the counter; we could keep those in some separate
memo, or perhaps store them in the nodes themselves. The counter is only
useful if a region has multiple parents, so we know when we've visited them
all; in this case, as soon as the counter goes from $0$ to $1$, we're done.
Also, storing $z$ is mostly useful if we're not going to use it immediately,
and so it may be a bit wasted in this case.

Be that as it may, once the counter hits the threshold $m$ (the number of
parents of the region), we visit the region node. Here we store the radius
$r$, but also one or more coefficients in a vector $a$. Note that $m$ here is
the number of entries in $a$, stored as part of the vector (or implicit, if
the length is fixed). In this case, $m=1$, $a=[1]$ and $f(x) = ax = x$. That
means the overlap check reduces to that of the BS-tree:
\[
    f(z) \leq r + f(s) \iff  az \leq r + as \iff z \leq r + s
\]
There is no magic in the use of two children here; we may very well increase
this number, as in the M-tree, for example~\cite{Ciaccia:1997}.
(The M-tree adds another twist, which we'll return to in
\cref{sec:intersections}.)

There's also the VP-tree~\cite{Uhlmann:1991,Yianilos:1993} and its relatives
such as LC~\cite{Chavez:2005}, where there's a single ball that separates the
inside from the outside. In that case, we get a different transformation:
\bigskip
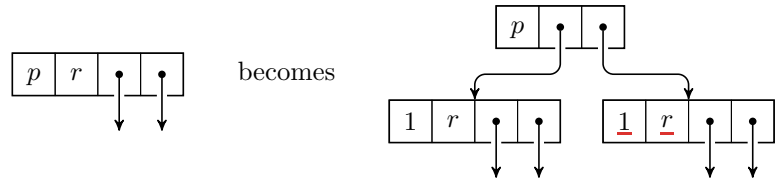
\begin{center}
\begin{tikzpicture}[baseline=-.5ex]

\matrix[memblock] (BS) { p \& r \& \& \\};

\draw
    (BS-1-3) node[tail] (BS-left) {}
    (BS-1-4) node[tail] (BS-right) {}
;

\draw[pointer] (BS-left) -- +(0,-.75);
\draw[pointer] (BS-right) -- +(0,-.75);

\end{tikzpicture}%
\qquad becomes\qquad%
\begin{tikzpicture}[baseline=-.5cm+.125cm-.5ex]

\matrix[memblock, anchor=west] (P) at (2.5*16pt,.25) { p \& \& \\};

\matrix[memblock, anchor=west] (R) at (0,-1) {
|[text depth=0]| 1 \& r \& \& \\};

\matrix[memblock, anchor=west] (R2) at (5*16pt,-1) {
|[text depth=0]| \ngt{1}
\&
|[text depth=0]|
\ngt*{r}
\& \& \\};

\draw
    (P-1-2) node[tail] (P-left) {}
    (P-1-3) node[tail] (P-right) {}
    (R-1-3) node[tail] (R-left) {}
    (R-1-4) node[tail] (R-right) {}
    (R2-1-3) node[tail] (R2-left) {}
    (R2-1-4) node[tail] (R2-right) {}
;

\draw[pointer] (P-left) |- ($(P-1-2)!.5!(R)$) -| (R);
\draw[pointer] (P-right) |- ($(P-1-3)!.5!(R2)$) -| (R2);
\draw[pointer] (R-left) -- +(0,-.75);
\draw[pointer] (R-right) -- +(0,-.75);
\draw[pointer] (R2-left) -- +(0,-.75);
\draw[pointer] (R2-right) -- +(0,-.75);

\end{tikzpicture}
\end{center}
\bigskip

\noindent
The idea here is that the center point $p$ is shared between the ball (left
subtree) and its complement, the outside (right subtree). The only difference
between the two region nodes is that the outside one has its coefficient and
radius negated.\footnote{Here \ngt{x} is a space-saving shorthand for
$-x$.} At this point, a slight revision is in order. We have previously
assumed that $f$ is nondecreasing, i.e., that $a\geq 0$. That is no longer the
case! The more general version of the overlap check then uses $|a|s$, rather
than $as$.  What happens, then, is that the overlap criterion for the left
subtree is still $z \leq r + s$, but for the right one, we get:
\[
    az \leq -r + |a|s \iff -z \leq -r + s \iff z \geq r - s
\]
This is exactly as in the VP-tree, except that the surface of the ball is
included both for the inside \emph{and} the outside; we'd really like $z > r -
s$. This is a detail not handled by the framework (though it easily could be
amended to); however, it could only (presumably in rare cases) lead to false
positives, i.e., exploring subtrees unnecessarily, which won't produce any
wrong results. However, except for the goal of emulating the VP-tree, there is
no need to use the same radius in both regions. One could use $r_1$ and
$-r_2$, for example, and adapt each to cover only the points in each subtree.

\section{Intersections}
\label{sec:intersections}

In \cref{sec:balltrees}, our regions were individual balls and their
complements.\footnote{Strictly speaking, the closure of their complements, as
we don't use strict inequalities.} We can combine these two kinds of regions
to create \emph{shell} regions, by turning $a$ and $r$ into column vectors:
\[
    a = \begin{bmatrix}
        -1 \\
        \phantom{-}1
    \end{bmatrix}
    \qquad
    r =
    \begin{bmatrix}
        -r\mathrlap{'}\phantom{''}\\
        \phantom{-}r''\\
    \end{bmatrix}
\]
This gives a shell region around the single parent point $p$, as follows:
\bigskip
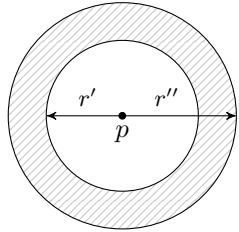
\begin{center}
\begin{tikzpicture}
    \def\RI{1}
    \def\RO{1.5}
    \draw[pattern={north east lines}, pattern color=black!20]
        (0,0) circle[radius=\RO]
        ;
    \draw[fill=white]
        (0,0) circle[radius=\RI]
        ;
    \draw[-stealth', shorten >=.5\pgflinewidth]%
        (0,0) -- (\RO,0);
    \draw (.5*\RI,0) node[above, font=\footnotesize] {$r\mathrlap{''}$};
    \draw[-stealth', shorten >=.5\pgflinewidth] (0,0) -- (-\RI,0);
    \draw (-.5*\RI,0) node[above, font=\footnotesize] {$r\mathrlap{'}$};
    \draw (0,0) node[point node] (p) {} node[below] {$p$};
\end{tikzpicture}
\end{center}
\bigskip
The membership check for a point $u$ with distance $x=\delta(p,u)$ is still
$ax\leq r$, but in this case, that means:
\begin{IEEEeqnarray*}{rCl}
    -x &\leq& -r' \\
    x &\leq& \phantom{-}r''
\end{IEEEeqnarray*}
This is, of course, equivalent to $r'\leq x \leq r''$. For the overlap check,
we take the absolute value for each row separately, so we still have $az\leq r
+ s$, which becomes (with some simplification):
\begin{IEEEeqnarray*}{rCl}
    z + s&\geq& r' \\
    z - s&\leq&  r''
\end{IEEEeqnarray*}
That is, $q$ must be so far away ($z$) that the $s$-ball around it reaches the
inside radius ($r'$) but not so far away that it ends up beyond the outside
radius ($r''$).

A classic metric index---the Burkhard--Keller tree---branches out using
multiple shells around a single center~\cite{Burkhard:1973}. In this case,
we'd simply use multiple shell regions, all with the same parent point.

There's not much point in using more than two rows when we have a single
focus, i.e., a center, as we'll only end up with a single ball, inverted ball
or shell, anyway. However, if we have more than one focus, we can add multiple
\emph{columns} to represent the intersection of multiple shells with
\emph{different} centers, yielding a coefficient matrix $A$.
For simplicity, let's say we wish to represent the intersection of two balls,
with respective centers $p_1$ and $p_2$. We use those points as the region's
parents, and region membership becomes $Ax\leq r$, with coefficients and radii
as follows:
\[
    A = \begin{bmatrix}
        1 & 0\\
        0 & 1
    \end{bmatrix}
    \qquad
    r =
    \begin{bmatrix}
        r_1\\
        r_2\\
    \end{bmatrix}
\]
The intersection of multiple shells has been used in, e.g.,
Brin's GNAT~\cite{Brin:1995} and its descendants, as well
as the PM-tree family of structures~\cite{Skopal:2004} (see also
\cref{sec:elimination}), and was later dubbed a \emph{cut region}
by Loko{\v{c}} et al.~\cite{Lokoc:2014}.

The M-tree combines balls and shells in an interesting way. Before even
computing $\delta(q,u)$ to perform the overlap check $\delta(q,u) \leq r + s$,
it executes a preliminary filtering step, with the check
\[
    |\delta(q,p) - \delta(p,u)| \leq r + s\mathrlap{\,,}
\]
or, with our established notation, $|z - x|\leq r + s$. The intuition here is
that $|z-x|$ is a lower bound for $d(q,u)$, a fact used in the standard pivot
filtering check $|z-x|\leq s$ (see \cref{sec:elimination}). Here, however, it's
plugged in as a lower bound in our \emph{ball} overlap check (with $u$ as our
ball center), creating a weakened, preliminary version.
This might seem like it requires introducing some new concept or indirection,
but that is not so. The check is still linear and is equivalent to a standard
shell region. This is easily seen by rewriting the check as follows:
\begin{IEEEeqnarray*}{rCl}
  - z + x &\leq& r + s \\
    z - x &\leq& r + s
\end{IEEEeqnarray*}
We can rewrite this to match our previous shell overlap check:
\begin{IEEEeqnarray*}{rCl}
   z + s &\geq& x - r \\
    z - s &\leq& x + r
\end{IEEEeqnarray*}
In other words, we here simply have a shell region with inner radius $x-r$ and
outer radius $x+r$, corresponding to our knowledge of the $r$-ball around $u$
before computing $\delta(q,u)$:
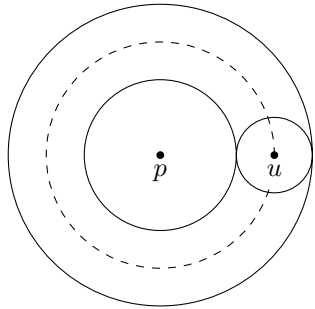
\begin{center}
\begin{tikzpicture}

    \draw
        (0,0) circle[radius=1]
        (0,0) circle[radius=2]
        (1.5,0) circle[radius=.5]
    ;
    \draw[dashed] (0,0) circle[radius=1.5];

    \draw
        (0,0) node[point node] {} node[below] {$p$}
        (1.5,0) node {} node[below, fill=white] {$u$}
        (1.5,0) node[point node] {}
        ;
\end{tikzpicture}
\end{center}
It would seem like we now have to store additional distances. Rather than just
keeping $x$ and $r$, we need to store $r$, $x-r$ and $x+r$. But is that really
so? Given our M-tree to sprawl translation, each point node is now the center
of multiple (quite possibly overlapping) shell regions, as well as a single
ball region enclosing them all. The only reason to keep this ball region is if
its radius is lower than the greatest radius of the shells. If we stuck
rigidly to our translation, this could happen---but if we simply kept our
shells as tight as possible around the subtrees, it could not. We then end up
storing just two distances per subtree, once more, and have a structure with a
behavior quite similar to, and no worse than, the M-tree.

\section{Elimination}
\label{sec:elimination}

The most common purpose of a pointer in an index structure is to lead you
toward further data to explore. There is a certain genre of structures,
however, that do the exact opposite---where instead of discovering data, you
\emph{eliminate} it. Take, for example, the LAESA structure~\cite{Mico:1994}:
a table of distances between so-called \emph{pivots} and the other points in
the data set. The query is compared to each of these pivots, and the computed
query--pivot distances, along with the stored pivot--data-point distances, are
used to determine whether any given data point may possibly be relevant. In
sprawl terms, each pivot--data-point distance represents a region:

\bigskip
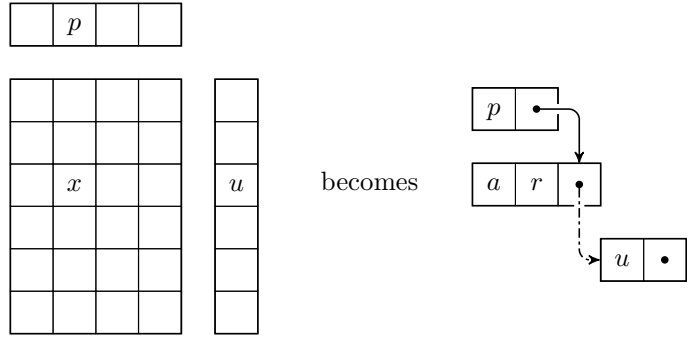
\begin{center}
\begin{tikzpicture}[baseline=-16pt * 3 + 8pt - .5ex]
\matrix[anchor=south east, memblock] at (0,12pt) {
    \& p \& \& \\
};
\matrix[anchor=north east, memblock] {
    \& \& \& \\
    \& \& \& \\
    \& x \& \& \\
    \& \& \& \\
    \& \& \& \\
    \& \& \& \\
};
\matrix[anchor=north west, memblock] at (12pt,0) {
    \\
    \\
    u \\
    \\
    \\
    \\
};
\end{tikzpicture}
\qquad becomes\qquad%
\begin{tikzpicture}[baseline=-1cm-.5ex]

\matrix[memblock, anchor=west] (P) { p \& \\};
\matrix[memblock, anchor=west] (U) at (3*16pt,-2) { u \& \\};

\matrix[memblock, anchor=west] (R) at (0,-1) {
a \&
r \&
\\};

\draw
    (P-1-2) node[tail] (P-ptr) {}
    (R-1-3) node[tail] (R-ptr) {}
    (U-1-2) node[tail] (U-ptr) {}
;

\draw[pointer] (P-ptr) -| (R-1-3);
\draw[pointer, dashdotted] (R-ptr) |- (U);

\end{tikzpicture}
\end{center}
\bigskip

\noindent
In this case, the region is a \emph{sphere}, a shell of width zero (i.e., with
identical inner and outer radii):
\[
    a = \begin{bmatrix}
        \phantom{-}1 \\
        -1
    \end{bmatrix}
    \qquad
    r = \begin{bmatrix}
        \phantom{-} x \\
        -x
    \end{bmatrix}
\]
As before, our overlap check is $az\leq r + s$, or:
\begin{IEEEeqnarray*}{rCr}
    z &\leq& x + s \\
    -z &\leq& -x + s
\end{IEEEeqnarray*}
Combined, this is the standard pivoting bound, $s\geq |x-z|$.
Now, however, we get to the more interesting point. The dotted pointer
indicates that we've got a potential \emph{elimination} on our hands. That is,
rather than saying ``if there's overlap, let's look at $u$, otherwise, let's
ignore it'' we turn it around, and say ``if there's overlap, let's ignore it,
otherwise, let's eliminate it.'' In the terminology of our earlier pseudocode,
that essentially means setting $u.\id{color}$ to \cn{black}. The exact
implementation here could be done in several ways. One could have different
region node types, for positive and negative regions (leading to discovery and
elimination, respectively), or have separate lists of positive and negative
child-edges, so the same region could both discover and eliminate points.
These are possible optimizations, but they don't substantially change the
behavior of the search.

It's possible to combine discovery and elimination, such as in the PM-tree. A
simplified version would consist of a ball tree, such as in
\cref{sec:balltrees}, along with a set of pivots with eliminating regions
around the subtrees. Specifically, the PM-tree uses shells around each
subtree, with global pivots, yielding something like the following:

\bigskip
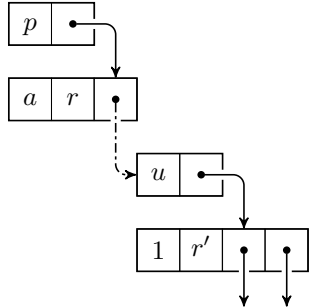
\begin{center}
\begin{tikzpicture}[baseline=-1cm-.5ex]

\matrix[memblock, anchor=west] (P) { p \& \\};
\matrix[memblock, anchor=west] (U) at (3*16pt,-2) { u \& \\};
\matrix[memblock, anchor=west] (R2) at (3*16pt,-3)
{ |[text depth=0]| 1 \& r' \& \& \\};

\matrix[memblock, anchor=west] (R) at (0,-1) {
a \&
r \&
\\};

\draw
    (P-1-2) node[tail] (P-ptr) {}
    (R-1-3) node[tail] (R-ptr) {}
    (U-1-2) node[tail] (U-ptr) {}
    (R2-1-3) node[tail] (R2-left) {}
    (R2-1-4) node[tail] (R2-right) {}
;

\draw[pointer] (P-ptr) -| (R-1-3);
\draw[pointer, dashdotted] (R-ptr) |- (U);

\draw[pointer] (U-ptr) -| (R2-1-3);
\draw[pointer] (R2-left) -- +(0,-.75);
\draw[pointer] (R2-right) -- +(0,-.75);

\end{tikzpicture}
\end{center}
\bigskip

An important thing to note about elimination is that it may be performed
\emph{lazily}. That is, we need not check for overlap with the various shell
regions associated with the shared, global points in the PM-tree until we've
established that we intersect the ball region in the tree itself. This kind of
laziness could be implemented by having pointers in the reverse direction,
without a need for counter updating. When considering a point, we would simply
look at the region parents and see if they had been examined yet (i.e., if
they were colored black).

It's possible to implement such things in different ways, of course; one
could, for example, have some \emph{parents} of a region be lazy, explored on
demand, or the like. Having such a mechanism, one could simply use the global
pivots of a PM-tree as lazy parents of every region in the tree, turning
them from balls into cut regions, removing the need for elimination
altogether.

The elimination perspective in LAESA could similarly be turned on its head, if
instead of multiple regions, we use a single region for each point, with all
pivots as its parents. This region would then be the intersection of all the
spheres, and a point would simply not be discovered if there were no
intersection.

This does not mean that we can do entirely without elimination, however. In
any scenario where we at one time are able to traverse a point, and at a later
time are not, this is the result of elimination. To my knowledge, the only
current structure where it is truly needed, even if one were to introduce
various forms of laziness optimization, is the AESA family of
indices~\cite{Vidal:1986}, where all points are available initially, and the
set of candidates is gradually whittled down. The order of traversal then
becomes crucial, as discussed in the next section.

\section{Priority}
\label{sec:priority}

The AESA family of indexing methods are all based on the same simple data
structure: a complete distance matrix between the data
points.\footnote{Because of symmetry, one need only store half of it, of
course.} The points are explored one by one, and at every step we eliminate
any of the remaining points we're able to. The elimination works just as in
LAESA; the difference is that the pivots aren't kept separate from the objects.
Rather than simply examining all the pivots in an arbitrary order, we now need
to be quite careful about which object to examine next, to minimize the number
of objects explored overall. A ubiquitous simplification here is to only focus
on the elimination power of the next object and select the one that will give
us the most bang for our buck.

We don't know which one that is, though. Rather, we must perform this choice
\emph{heuristically}, based on the information gathered so far, i.e., the
distances from each examined point to the query and to the candidates for
examination. If we represent these distances by vectors $z$ and $x$, the
original AESA
used $\|z-x\|_1$ while a revised version used $\|z-x\|_\infty$,
simply choosing the object whose filtering lower bound is the smallest, i.e.,
the one that's furthest away from being eliminated.
Later, there was iAESA~\cite{Figueroa:2010}, which instead used Spearman's
footrule between permutations of the previously examined objects, sorted by
distance to the query and the candidate.
Even more recently, Socorro et al.\@ introduced the two-phase PiAESA
method~\cite{Socorro:2011},
which initially uses a set of preselected pivots (like LAESA),
chosen for their general filtering power; once enough objects have been
explored, it switches to the classic AESA behavior. Many variations are
possible here, of course; for example, one might use regression or machine
learning to estimate distances or filtering power or the
like~\cite{Edsberg:2010,Murakami:2013}.

From a sprawl-of-ambits point of view, these methods are essentially the same:
A complete directed graph of elimination edges, where each edge has a single
sphere region. The priority or heuristic used to select the next available
point is left unspecified. What \emph{is} relevant, however, is when and how
to compute or update the heuristic. In the simplest, most naive
implementation, on might merely iterate over all available objects in each
step, computing an arbitrary black-box priority for each, based on the
knowledge gathered so far. It's possible, however, to let priority updates
piggyback on other traversal operations.

For example, if the heuristic is based on how hard a point is to get rid of,
one might update the priority every time the point is rediscovered and every
time one fails to eliminate it. In each of these cases, a lower bound on the
distance is computed, and one may then simply keep the sum or maximum, as in
AESA.

For structures without elimination, such as the majority of search trees,
priority is not relevant to the number of distance computations needed to
resolve a range query; the behavior will be the same, regardless of the
ordering. For $k$NN queries, however, priority can be crucial, as the covering
radius of the result set tends to shrink as good candidates are found, and
this will improve the chances of eliminating subtrees.

\section{Non-trees}
\label{sec:nontrees}

Index structures tend to be tree-shaped, more or less, especially if we ignore
the eliminating parts. One early exception is the \emph{excluded middle
vantage point forest} introduced by Yianilos~\cite{Yianilos:1999}. This
structure is still \emph{mostly} tree-shaped---or, as the name implies,
forest-shaped. That is, it primarily consists of a collection of trees.
However, these trees are connected to each other, making the whole thing a
directed, acyclic graph.

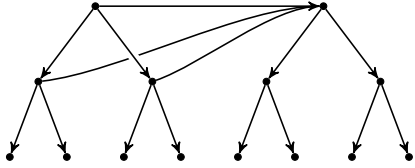
\begin{center}
\begin{tikzpicture}[xscale=.75]
\foreach \shift/\side in {{(0,0)}/L,{(4,0)}/R} {
    \draw[shift=\shift]
        (+0.0,+0) node[point node] (\side-1) {}
        (-1.0,-1) node[point node] (\side-2) {}
        (+1.0,-1) node[point node] (\side-3) {}
        (-1.5,-2) node[point node] (\side-4) {}
        (-0.5,-2) node[point node] (\side-5) {}
        (+0.5,-2) node[point node] (\side-6) {}
        (+1.5,-2) node[point node] (\side-7) {}
        ;
}
\draw[semithick]
    (L-1) edge[-stealth'] (R-1)
    (L-2) edge[looseness=.7, out=5, in=180] (R-1)
    (L-3) edge[looseness=.7, out=15, in=180] (R-1)
;
\draw[semithick, -stealth']
    (L-1) edge (L-2) edge[over] (L-3)
    (L-2) edge (L-4) edge (L-5)
    (L-3) edge (L-6) edge (L-7)
;
\draw[semithick, -stealth']
    (R-1) edge (R-2) edge (R-3)
    (R-2) edge (R-4) edge (R-5)
    (R-3) edge (R-6) edge (R-7)
;
\end{tikzpicture}
\end{center}

\noindent
The trees are essentially VP-trees with three regions rather than two: an
inner ball, a middle shell, and an outer inverted ball. For queries whose
radius is less than half the width of the middle shell, the search will never
traverse more than one of the inner and outer subtrees---a major selling-point
of the structure. There may still be points located in the separating shell,
though, and these must also be indexed!

The idea is to gather up all the points that end up in any separating shell
throughout the tree, and build a \emph{new} tree from those, in the same
manner (possibly leading to a third tree, and so on). We then simply make the
root of this new tree the single child of every shell region in the first
tree, as in the following, where
\(
    a = [
        -1\ 1
        ]^t
\)
and
\(
    r = [
        -r'\ r''
        ]^t
\):

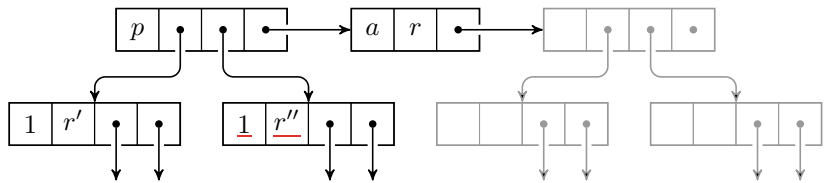
\NewDocumentCommand \VPRoot {mmmmmm} {

    \matrix[memblock, anchor=west] (#6P) at (2.5*16pt,.25) { #1 \& \& \& \\};

    \matrix[memblock, anchor=west] (#6R) at (0,-1) {
    |[text depth=0]| #2 \& #3 \& \& \\};

    \matrix[memblock, anchor=west] (#6R2) at (5*16pt,-1) {
    |[text depth=0]| #4
    \&
    |[text depth=0]|
    #5
    \& \& \\};

    \draw
        (#6P-1-2) node[tail] (#6P-left) {}
        (#6P-1-3) node[tail] (#6P-right) {}
        (#6P-1-4) node[tail] (#6P-next) {}
        (#6R-1-3) node[tail] (#6R-left) {}
        (#6R-1-4) node[tail] (#6R-right) {}
        (#6R2-1-3) node[tail] (#6R2-left) {}
        (#6R2-1-4) node[tail] (#6R2-right) {}
    ;

    \draw[pointer] (#6P-left) |- ($(#6P-1-2)!.5!(#6R)$) -| (#6R);
    \draw[pointer] (#6P-right) |- ($(#6P-1-3)!.5!(#6R2)$) -| (#6R2);
    \draw[pointer] (#6R-left) -- +(0,-.75);
    \draw[pointer] (#6R-right) -- +(0,-.75);
    \draw[pointer] (#6R2-left) -- +(0,-.75);
    \draw[pointer] (#6R2-right) -- +(0,-.75);

}

\begin{center}
\begin{tikzpicture}[baseline=-.5cm+.125cm-.5ex]

\VPRoot{p}{1}{r'}{\ngt{1}}{\ngt*{r''}}{}
\begin{scope}[xshift=10*16pt, draw=secondary color, tail/.append
    style={fill=secondary color}]
\VPRoot{}{}{}{}{}{R}
\end{scope}

\matrix[memblock, anchor=west]
    (T) at (8*16pt,0.25) {a \& r \& \\};
\draw (T-1-3) node[tail] (T-child) {};

\draw[pointer] (P-next) -- (T);
\draw[pointer] (T-child) -- (RP);

\end{tikzpicture}
\end{center}

\noindent
An essentially equivalent structure, at least from a bird's-eye view, is the
D-index~\cite{Dohnal:2003}. There, too, we have a multitude of shell regions
separating inner and outer subsets, with the shells leading to a secondary
structure, and so on. The main difference is that where the excluded middle
vantage point forest uses tree traversal to determine which intersection of
inner and outer regions a given point falls into, with the centers found along
the path from the root, the D-index provides a fixed set of shared centers
from the beginning, in a manner similar to the so-called fixed-queries
tree~\cite{Baeza:1994}. Several levels are, in essence, collapsed, and the
correct subtrees or leaves, representing the intersection of multiple shell or
ball regions, are found directly, using hashing. This is an optimization that
does not affect the high-level behavior (i.e., which points are examined).

\section{Hyperplanes}

In \cref{sec:elimination}, we created sphere and shell regions by having two
radii, and thus two rows in our coefficient matrix, ending up with a column
vector $[1\ {-1}]^t$. But we could also just use a row vector $a=[1\ {-1}]$,
along with a single radius. This means we need two parents, or \emph{foci},
$p_1$ and $p_2$, and we finally get a pivot vector $z=[z_1\ z_2]^t$.
The overlap check becomes:
\[
    az \leq r + \|a\|_1s
\]
Here $\|a\|_1$ is the sum of absolute values. If we use $r=0$, this
corresponds to a metric half-space, separated by a midset or hyperplane. The
overlap check then simplifies to the standard one~%
\cite{Uhlmann:1991}:
\[
    z_1-z_2 \leq 2s
\]
We are here defining the region of points closer to $p_1$ than $p_2$. If we
wish to have \emph{multiple} contrasting objects, modeling general Voronoi
cells or Dirichlet domains~%
\cite{Navarro:2002}, we can just add parent points, as well as some rows and
columns. Let's say, for example, we wish to describe the region of points that
are closer to $p_1$ than both $p_2$ and $p_3$. We'd then use all three as
parents of our region, and use the following coefficients and radii:
\[
    A =
    \begin{bmatrix}
        1 & -1 & \phantom{-}0 \\
        1 & \phantom{-}0 & -1
    \end{bmatrix}
    \qquad
    r =
    \begin{bmatrix}
        0 \\
        0
    \end{bmatrix}
\]
This corresponds to the following overlap check, where both inequalities must
hold for there to be overlap:
\begin{IEEEeqnarray*}{rCl}
    z_1-z_2 \leq 2s \\
    z_1-z_3 \leq 2s
\end{IEEEeqnarray*}
One may extend this to an arbitrary number of foci in the obvious manner.

\section{Other Conics}

The hyperplane case is easy enough to extend to (generalized)
ellipses~%
\cite{Uhlmann:1991,Dohnal:2001}, by using coefficients $a =
[1\ 1]$ and the appropriate radius, yielding the following overlap check:
\[
    az \leq r + \|a\|_1s \iff z_1 + z_2 \leq r + 2s
\]
Or we can get shifted half-spaces, what amounts to metric
hyperbolas~%
\cite{Dohnal:2001,Lokoc:2010}, by adjusting the radius away from $0$. That
is, we still have $a=[1\ {-1}]$ but we have $r\neq 0$, yielding the following
slightly more general check:
\[
    z_1-z_2\leq r + 2s
\]
This, then, represents not the points that are closer to $z_1$ than to $z_2$,
but where the distances differ by a given value (i.e., the radius). That is,
membership for a point with distance vector $x$ is $ax\leq r$, i.e., $x_1 -
x_2 \leq r$.

\section{Other Queries}

Nearest neighbor queries ($k$NN) have been mentioned briefly already. A
general approach is to maintain the (up to) $k$ points closest to $q$ found so
far, letting the search radius $s$ be an upper bound on the distance to
the $k$th nearest neighbor. Beyond updating $s$ during the search, the
procedure is the same.

A generalization that does not seem to have been explored is using other
regions than balls as queries. After all, if a ball query
works well in a tree built from hyperplanes, there's nothing stopping us from
using a hyperplane query in a tree built from balls. That is, we might have a
prototypical example object $q$, and a prototypical \emph{counter}-example
$q'$, and we then search our index for objects closer to $q$ than $q'$.
(Such queries were briefly mentioned by Uhlmann~\cite{Uhlmann:1991}.)
Or maybe we have two prototypes, and wish to find the $k$ objects with the
lowest average distance to $q$ and $q'$, resulting in an ellipsoid query.

More generally, our query might consist of a \emph{weighted combination} of
query objects, looking for points with a low weighted sum of query distances.
In other words, we may use an arbitrary linear ambit as our
query~%
\cite[\S\,3.2.1]{Hetland:2019}. As long as the ambit coefficients of the
query, \emph{or} those in the tree, are all non-negative, determining
query--region overlap is straightforward, as we shall see.

\section{\dots\,and Beyond}

It ought to be quite clear that the sprawls and ambits used so far have been
quite limited. The sprawls have mostly been tree-like, and the coefficients of
the ambits have been $1$ or $-1$, with at most two nonzero coefficients to a
row. Countless variations are possible, both in how the sprawls are put
together and in how the ambits are parameterized.

Determining whether an arbitrarily constructed sprawl is correct is a hard
problem~\cite[Thm.\,2.3.2]{Hetland:2019}. However, extrapolating from
existing index structures, we may quite easily ensure that the sprawls we
construct are \emph{responsible}, in which case they are guaranteed to be
correct. Roughly, responsibility means that for every point $p$, there is a
set of edges we can traverse that will lead us to it, and that the regions of
those edges contain $p$, as do the regions of any negative edges that might
disrupt that traversal. For the case where the positive edges of our structure
are acyclic, this can be dealt with locally, where the responsibilities of a
node's incoming edges depend only on those of the outgoing
ones~\cite[Obs.\,2.3.10]{Hetland:2019}. Thus it ought to be possible to mix
and match quite freely, perhaps even using heuristic search to look for
efficient structures automatically.

As for regions, any coefficient matrix and radius vector yields a valid linear
ambit, usable as a region or a query. For a query ambit $Q$ with coefficient
vector $c$ and radius $s$, and a region ambit $R$ with coefficient vector $a$
and radius $r$, with $a$ or $c$ non-negative and $\|a\|_1,\|c\|_1=1$, if $R$
and $Q$ intersect, then
\begin{equation}
    r + s \geq a \mkern1mu Z \mkern1mu c^t\eqcomma
\end{equation}
where $z_{ij}$ is the distance between focus $p_i$ of $R$ and focus $q_j$ of
$Q$~\cite[Thm\,3.1.2]{Hetland:2019}.
With this overlap check, one can use ambit queries with existing index
structures, and one could extend existing indexes with additional regions,
without adding any distance computations. In an tree structure where several
points are explored when deciding which subtrees to visit, arbitrary subsets
of these could be used to construct additional filtering predicates for any
subtrees, merely by adding radii and possibly coefficients.\footnote{This
approach has been tentatively explored by my students Carlsen and
Moe~\cite{Carlsen:2020}.}

Finally, one may go beyond the limits of linearity. For example, using any
(multi-parameter) non-decreasing \emph{metric-preserving} function $f$ to
calculate remoteness, we may still use the original overlap
check~\eqref{eq:overlap}~\cite[\S\,3.5]{Hetland:2019}.
This opens the door to a wide range of learning and optimization methods for
adapting regions to points in ways that improve search performance.

\def\doi#1{\href{http://doi.org/#1}{\nolinkurl{doi:#1}}}
\bibliography{paper}

\end{document}